\documentclass[preprintnumbers,superscriptaddress,nofootinbib,aps,prd,floatfix,twocolumn]{revtex4-2}
\pdfoutput=1
\usepackage{enumerate}
\usepackage{amsmath,amssymb}
\usepackage{graphicx}
\usepackage{slashed}
\usepackage{xspace,slashed}
\usepackage{float}
\usepackage[dvipsnames]{xcolor}
\usepackage{hyperref}
\hypersetup{colorlinks=true, citecolor=Blue, urlcolor=Blue, linkcolor=Blue}
\usepackage[normalem]{ulem}
\usepackage{subfigure,orcidlink}
\usepackage{multirow,array}
\usepackage{braket}
\usepackage{marginnote}
\usepackage{eurosym}
\usepackage{tabularray}
\UseTblrLibrary{booktabs}
\usepackage{bbding}
\usepackage{mathrsfs}

\begin{document}


\title{Probing the dynamical inverse seesaw mechanism\\with low-frequency gravitational waves}


\author{Debasish Borah \orcidlink{https://orcid.org/0000-0001-8375-282X}}
\email{dborah@iitg.ac.in}
\affiliation{Department of Physics, Indian Institute of Technology Guwahati, Assam 781039, India}

\author{Sounak Dutta \orcidlink{https://orcid.org/0009-0007-8179-5261}}
\email{d.sounak@iitg.ac.in}
\affiliation{Department of Physics, Indian Institute of Technology Guwahati, Assam 781039, India}

\author{Partha Kumar Paul \orcidlink{https://orcid.org/0000-0002-9107-5635}}
\email{ph22resch11012@iith.ac.in}
\affiliation{Department of Physics, Indian Institute of Technology Hyderabad, Kandi, Telangana 502285, India}

\author{Indrajit Saha \orcidlink{https://orcid.org/0000-0002-7459-0838}}
\email{s.indrajit@iitg.ac.in}
\affiliation{Department of Physics, Indian Institute of Technology Guwahati, Assam 781039, India}

\author{Narendra Sahu \orcidlink{https://orcid.org/0000-0002-9675-0484}}
\email{nsahu@phy.iith.ac.in}
\affiliation{Department of Physics, Indian Institute of Technology Hyderabad, Kandi, Telangana 502285, India}

\date{\today}
\begin{abstract}
We study the possibility of probing the dynamical inverse seesaw mechanism for the origin of light neutrino masses via the detection of stochastic gravitational waves (GW) in the low-frequency regime currently being probed by pulsar timing arrays. As the lepton number-violating term in inverse seesaw typically remains in the sub-MeV ballpark, its dynamical origin naturally brings the possibility of a low-scale first-order phase transition, which can be probed at low-frequency GW experiments. We also find interesting complementarity with heavy neutral lepton searches, as GW experiments remain sensitive to parameter space with small active-sterile mixing, which is out of reach for most particle physics experiments.
\end{abstract}	

\maketitle
\preprint{\href{https://doi.org/10.1103/jr3f-6g27}{Phys. Rev. D 114 (2026) 9 095038}}
	
\noindent
\section{Introduction}
The origin of light neutrino masses and mixing \cite{ParticleDataGroup:2024cfk} has been one of the longstanding problems in particle physics. The popular explanation arises within the framework of the seesaw mechanism, where heavy beyond standard model (BSM) degrees of freedom are responsible for generating a hierarchy between the scale of neutrino mass and the electroweak scale. While canonical seesaw models like type-I \cite{Minkowski:1977sc, Gell-Mann:1979ijt, GellMann:1980vs, Mohapatra:1979ia,Sawada:1979dis,Yanagida:1980xy, Schechter:1980gr}, type-II \cite{Mohapatra:1980yp, Schechter:1981cv, Wetterich:1981bx, Lazarides:1980nt, Brahmachari:1997cq} and type-III \cite{Foot:1988aq} typically correspond to new physics at very high scale, in the presence of additional approximate symmetries or textures, it is possible to have larger Yukawa couplings even with TeV scale seesaw \cite{Shaposhnikov:2006nn, Kersten:2007vk, Moffat:2017feq}, which keeps such scenarios within reach of future experiments \cite{Klaric:2020phc, Drewes:2021nqr, Hernandez:2022ivz,Abdullahi:2022jlv}. One of the most popular low-scale seesaw mechanisms is the inverse seesaw mechanism \cite{Mohapatra:1986aw, Mohapatra:1986bd, Gonzalez-Garcia:1988okv, Bernabeu:1987gr, Gavela:2009cd, Catano:2012kw}, which explains light neutrino masses by introducing a small lepton-number-violating term while keeping the heavy fermions at the TeV scale with large Yukawa couplings. As such, direct detection prospects remain absent for high scale seesaw scenarios. Several recent attempts also considered the possibility of probing them indirectly via stochastic gravitational wave (GW) observations \cite{Dror:2019syi, Blasi:2020wpy, Fornal:2020esl, Samanta:2020cdk, Barman:2022yos, Huang:2022vkf, Dasgupta:2022isg, Okada:2018xdh, Hasegawa:2019amx, Borah:2022cdx, Borah:2022vsu, Barman:2023fad, Borah:2023saq,Borah:2025bfa, Borah:2026kfo}. In these works, the sources of stochastic GW were considered to be cosmic strings \cite{Dror:2019syi, Blasi:2020wpy, Fornal:2020esl, Samanta:2020cdk, Borah:2022vsu}, domain walls \cite{Barman:2022yos, Barman:2023fad,Saikawa:2017hiv,Roshan:2024qnv,Bhattacharya:2023kws,Blasi:2022ayo,Blasi:2023sej, Borah:2024kfn,Paul:2024iie, Borboruah:2024lli, Borah:2025bfa, Borah:2026kfo} or bubbles generated at first order phase transition (FOPT) \cite{Huang:2022vkf, Dasgupta:2022isg, Okada:2018xdh, Hasegawa:2019amx, Borah:2022cdx, Borah:2023saq}. Interestingly, the hints for such stochastic GW have already been obtained from four different pulsar timing array (PTA) experiments, namely NANOGrav \cite{NANOGrav:2023gor}, European Pulsar Timing Array (EPTA) together with the Indian Pulsar Timing Array (InPTA) \cite{EPTA:2023fyk} and PPTA \cite{Reardon:2023gzh}, all part of the consortium called International Pulsar Timing Array (IPTA).

Motivated by this, we study the possibility of probing the inverse seesaw at PTA-based experiments. The low-scale nature of the inverse seesaw makes it natural to have signatures at low-frequency GW experiments like PTA. The origin of GW is related to the low-scale FOPT responsible for generating the inverse seesaw scale dynamically. An additional scalar singlet is responsible for driving the low-scale FOPT while also generating the lepton-number-violating terms of the inverse seesaw scenario dynamically. For the TeV scale inverse seesaw, the lepton number violating term lies in the $\lesssim \mathcal{O}(1)$ MeV, keeping the corresponding FOPT-generated GW in the nHz or PTA ballpark. We also discuss the interesting complementarity among different GW experiments \cite{NANOGrav:2023ctt,Garcia-Bellido:2021zgu,Garcia-Bellido:2021zgu,Sesana:2019vho,Weltman:2018zrl,NANOGrav:2023gor} and heavy neutral lepton (HNL) searches \cite{CMS:2018iaf,CMS:2018jxx,ATLAS:2019kpx,ATLAS:2015gtp,DELPHI:1996qcc,L3:1992xaz,L3:2001zfe,Belle:2013ytx,CHARM:1985nku,CHARMII:1994jjr,Bernardi:1987ek,Barouki:2022bkt,NA3:1986ahv,NA62:2020mcv,BESIII:2019oef,Blennow:2016jkn,delAguila:2008pw,deBlas:2013gla,Antusch:2014woa,Das:2017zjc,Borexino:2013bot,Hagner:1995bn,Derbin:1993wy,PIENU:2017wbj,T2K:2019jwa,Boyarsky:2009ix,Ruchayskiy:2012si} for our scenario, keeping the model accessible at a variety of experiments across energy, intensity, and cosmic frontiers. 
\section{Dynamical Inverse Seesaw}
The minimal inverse seesaw mechanism considers two different types of heavy singlet fermions $N_R, S_L$ with the relevant Lagrangian given by 
\begin{equation}
  -\mathcal{L} \supset Y_D \overline{\ell_L} \tilde{\Phi} N_R + M_R \overline{N_R} S_L + \frac{1}{2} \mu \overline{S^c_L} S_L + {\rm H.c.}  
\end{equation}
After the electroweak symmetry breaking, we can write down the full neutral fermion mass matrix in the $(\nu_L, N_R, S_L)$ basis as 
\begin{equation}
    M_\nu = \begin{pmatrix}
        0 & M_D & 0 \\
        M^T_D & 0 & M_R \\
        0 & M^T_R & \mu \\
    \end{pmatrix},
\end{equation}
where $M_D = Y_D v_{\rm ew}/\sqrt{2}$ with $v_{\rm ew} =246$ GeV being the vacuum expectation value (VEV) of the Standard Model (SM) Higgs, $\Phi$. In the limit $\mu \ll M_D \ll M_R$, the light $3\times 3$ neutrino mass matrix can be found after block-diagonalization as 
\begin{eqnarray}
    m_\nu &=& -M_D M^{-1}_R \mu (M^T_R)^{-1} M^T_D.
\end{eqnarray}
Thus, for $\mu \lesssim \mathcal{O}(\rm MeV)$, sub-eV light neutrino mass can be obtained with $M_D, M_R$ around a few tens of GeV and a few tens of TeV, respectively.

Even though a small lepton number violating $\mu$ term is technically natural \cite{tHooft:1979rat}, it is imposed by hand, lacking explanation behind its smallness compared to the electroweak scale. This has led to several works which provide a dynamical origin of inverse seesaw \cite{Bazzocchi:2010dt, Ma:2009gu, Dias:2012xp, DeRomeri:2017oxa, Baldes:2013eva, Panda:2022kbn, Georis:2025kzv, Berbig:2024uwm}. We adopt a minimal setup where a complex singlet scalar $\phi$ is introduced to generate the $\mu$ term dynamically, while a spectator complex scalar field $\phi'$ assists in the FOPT driven by $\phi$. A complete UV completion is given in Appendix \ref{appen1}. The relevant part of the Lagrangian responsible for the dynamical generation of the $\mu$ term is 
\begin{equation}
    -\mathcal{L} \supset \frac{1}{2} Y_\mu \phi \overline{S^c_L} S_L + {\rm H.c.}
    \label{eq:Ymu}
\end{equation}
The scalar potential involving $\phi, \phi'$ is given as
\begin{eqnarray}
    V(\phi,\phi^\prime)&=&-\mu_{\phi}^2\phi^\dagger\phi+\lambda_{\phi}(\phi^\dagger\phi)^2+\mu_{\phi^\prime}^2\phi^{\prime\dagger}\phi^{\prime}+\lambda_{\phi^\prime}(\phi^{\prime\dagger}\phi^{\prime})^2\nonumber\\&&+\frac{\kappa_\phi}{3}(\phi^3+(\phi^\dagger)^3)+\lambda_{\phi\phi^\prime}(\phi^\dagger\phi)(\phi^{\prime\dagger}\phi^{\prime}).
    \label{eq:vtree}
\end{eqnarray}
Writing the scalar fields as
\begin{eqnarray}
\phi=\frac{\phi+v+i\eta}{\sqrt{2}},\phi^\prime=\frac{\phi^\prime+i\eta^\prime}{\sqrt{2}},
\end{eqnarray}
the $\mu$ term can be obtained as $\mu = Y_\mu v/\sqrt{2}$. Since we generate the lepton number violating term of inverse seesaw dynamically, a low scale realization of this seesaw naturally requires a MeV scale symmetry breaking driven by the scalar field $\phi$. We study the possibility of a MeV scale first-order phase transition (FOPT) driven by the singlet scalar field $\phi$.

The scale of symmetry breaking or the $\mu$ term indirectly controls the heavy fermion mass $M_R$. This effectively controls the active-sterile mixing angle, which is parametrized as
\begin{eqnarray}
    \Theta\equiv M_DM_{R}^{-1}=U_{\rm PMNS}^*\sqrt{m_\nu^d}R (\sqrt{\mu})^{-1},
\end{eqnarray}
where $U_{\rm PMNS}$ is the PMNS lepton mixing matrix, $m_\nu^d$ is the diagonal light neutrino mass matrix with eigenvalues $m_1,m_2$, and $m_3$\footnote{For simplicity, we choose $m_1=0$ and $R$ to be an identity matrix for the analysis.}. The details of this parametrization are given in Appendix \ref{app:yukawa}.

\begin{figure}
    \centering
 \includegraphics[scale=0.26]{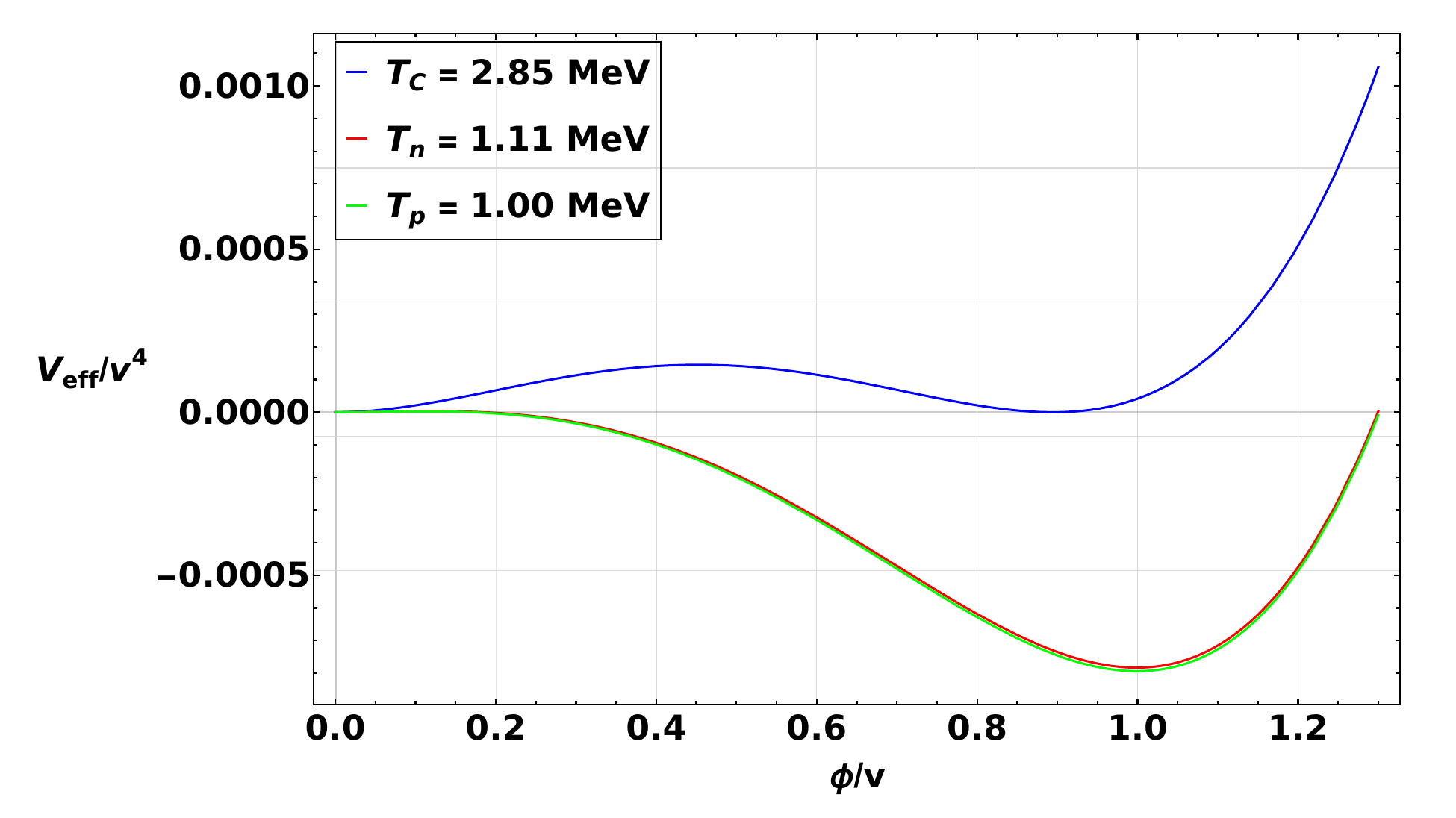}
    \caption{The effective potential as a function of the field value, evaluated at three cosmologically relevant temperatures for BP2: the critical temperature \(T_C\), at which the two minima are degenerate; the nucleation temperature \(T_n\), at which bubble nucleation becomes efficient; and the percolation temperature \(T_p\), at which the phase transition completes. We have appropriately used the value of the vacuum expectation value \(v\) to normalize both axes.}
    \label{fig:bp2}
\end{figure}
\begin{figure}
    \centering   \includegraphics[scale=0.24]{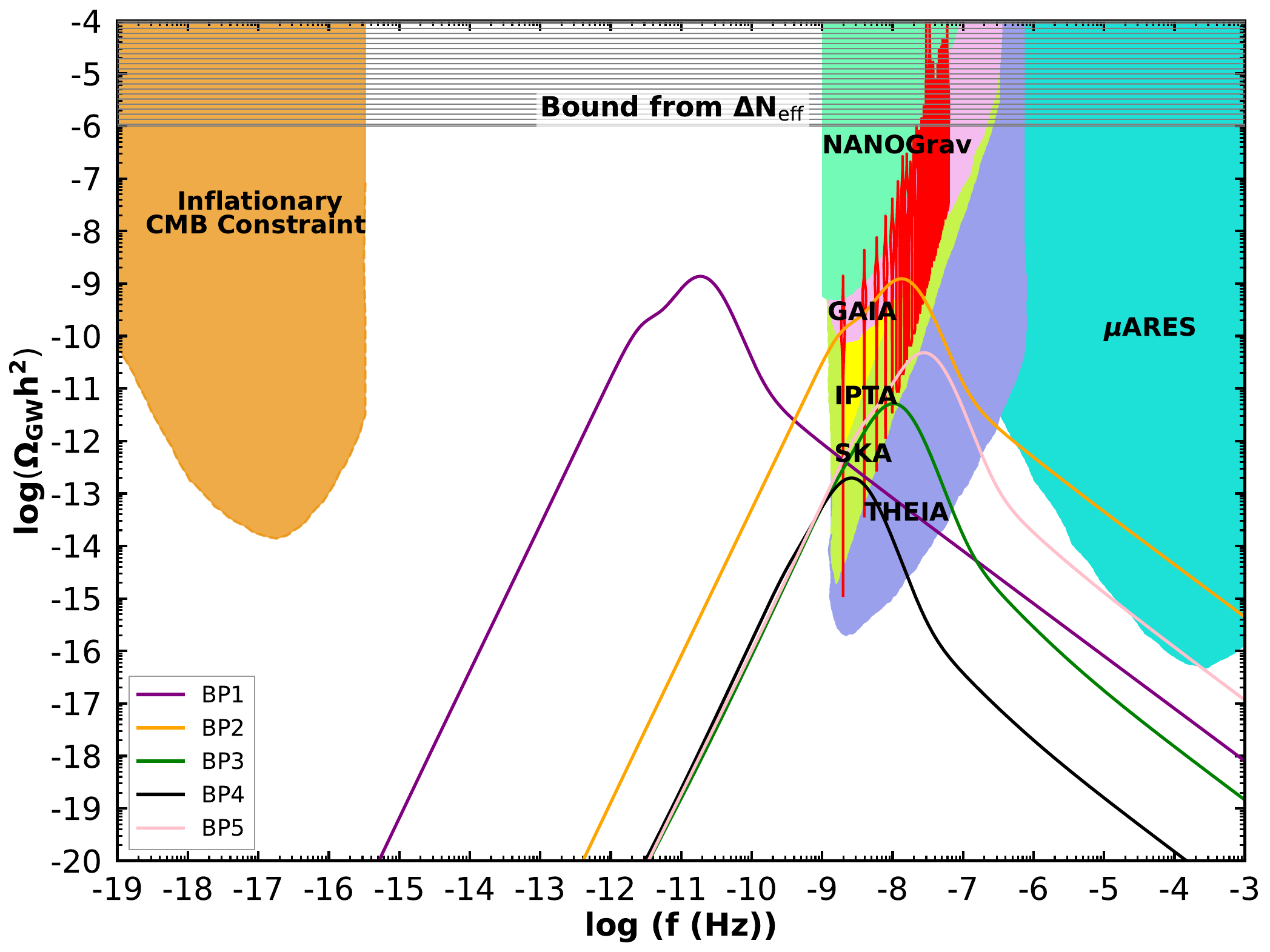}
    \caption{The gravitational wave energy density spectrum \(\Omega_\mathrm{GW}h^2\) as a function of frequency $f$, generated from the first-order phase transition of our model for the benchmark points listed in Table~\ref{tabgw}. The shaded regions indicate the projected sensitivity curves of current and future experiments: NANOGrav, GAIA, SKA, THEIA, the IPTA collaboration, and \(\mu\)ARES. The yellow shaded region marks the exclusion zone imposed by the inflationary CMB constraint, ruling out any GW signal in that parameter space. The hatched region at the top denotes the upper bound from \(\Delta N_{\rm eff}\).}
    \label{fig:gwspec}
\end{figure}
\begin{table*}
    \centering
    \begin{tabular}{|c|c|c|c|c|c|c|c|c|c|c|}
\hline
$\lambda_{\phi\phi^\prime}$ & $v$ & $\lambda_\phi$ &$\kappa_\phi$  &$T_C$ &$T_n$& $T_p$ &\rule{0pt}{12pt} $\alpha_p$ & $\beta/H_p$ & $T_\mathrm{RH}$ & Remarks \\\hline
 0.7 & 10 keV & 0.01 & -0.13 keV  & 2.86 keV& 1.28 keV& 1.18 keV & 3.58 & \rule{0pt}{12pt}101.78 & 1.71 keV &BP1\\\hline 
0.82& 10 MeV& 0.01&  -0.12 MeV& 2.85 MeV& 1.11 MeV& 1.01 MeV& \rule{0pt}{12pt}2.31& 76.15& 1.35 MeV& BP2\\\hline
 0.25& 2 MeV& 0.01 & -0.03 MeV& 0.69 MeV& 0.39 MeV& 0.37 MeV& 0.18& \rule{0pt}{12pt}207.59& 0.38 MeV&BP3\\ \hline
0.9 & 60 keV & 0.01 & -0.45 keV  & 19.92 keV& 16.08 keV& 15.88 keV & 0.21 & \rule{0pt}{12pt}1479.77 & 16.29 keV & BP4\\ \hline
0.99 & 5 MeV & 0.1 & -0.86 MeV & 1.83 MeV & 1.18 MeV & 1.12 MeV & 0.53 & \rule{0pt}{12pt} 172.47 & 1.22 MeV & BP5\\ \hline
    \end{tabular}
    \caption{Benchmark parameter sets for the model, corresponding to points in the parameter space that yield a first-order phase transition. We have fixed $\mu_{\phi^\prime}=0$ and $\lambda_{\phi^\prime}=10^{-5}$ for the analysis. The associated gravitational wave spectra for these benchmarks are shown in Fig.~\ref{fig:gwspec}.}
    \label{tabgw}
\end{table*}

\begin{figure*}
\centering
\includegraphics[width=0.49\textwidth]{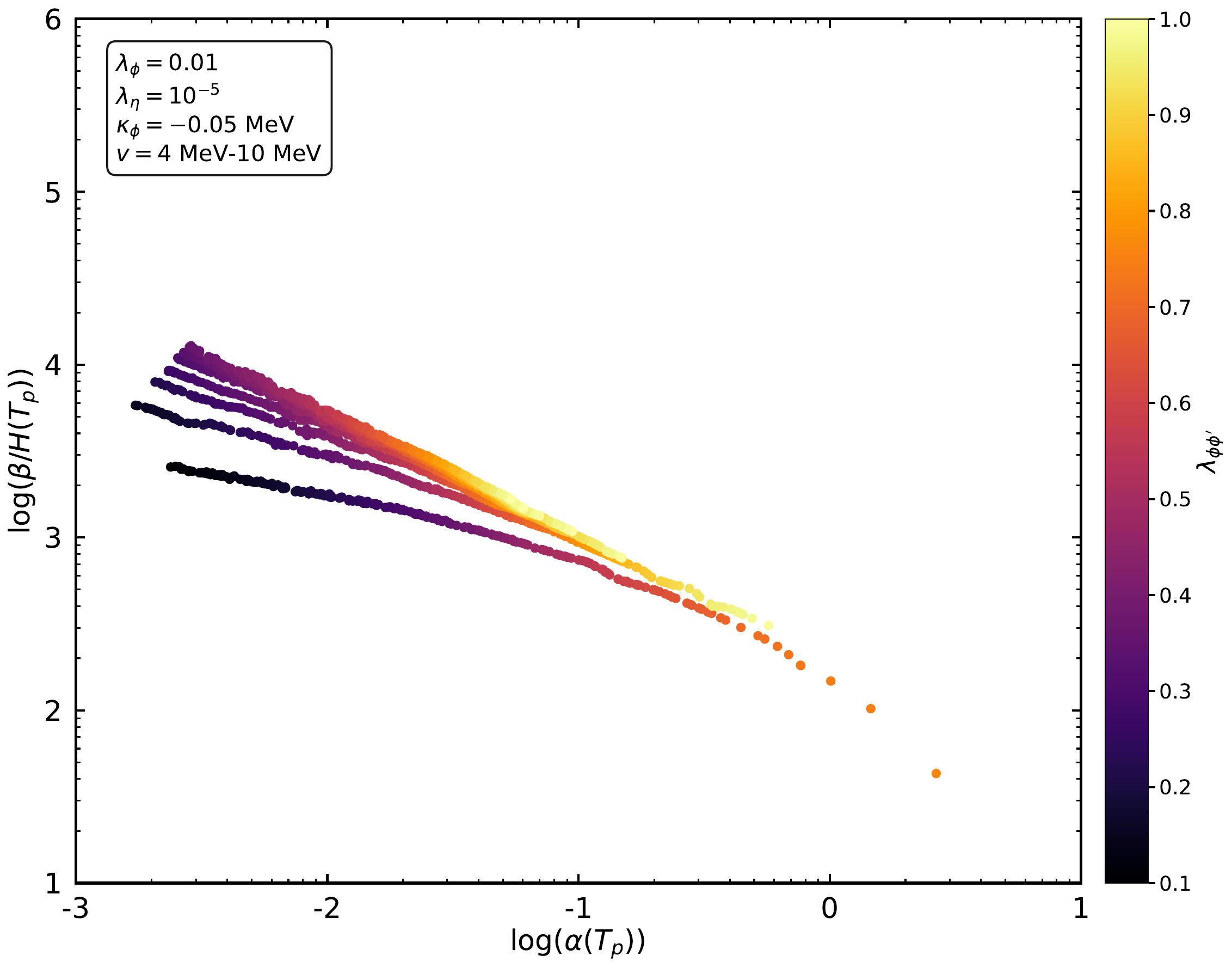}
\includegraphics[width=0.49\textwidth]{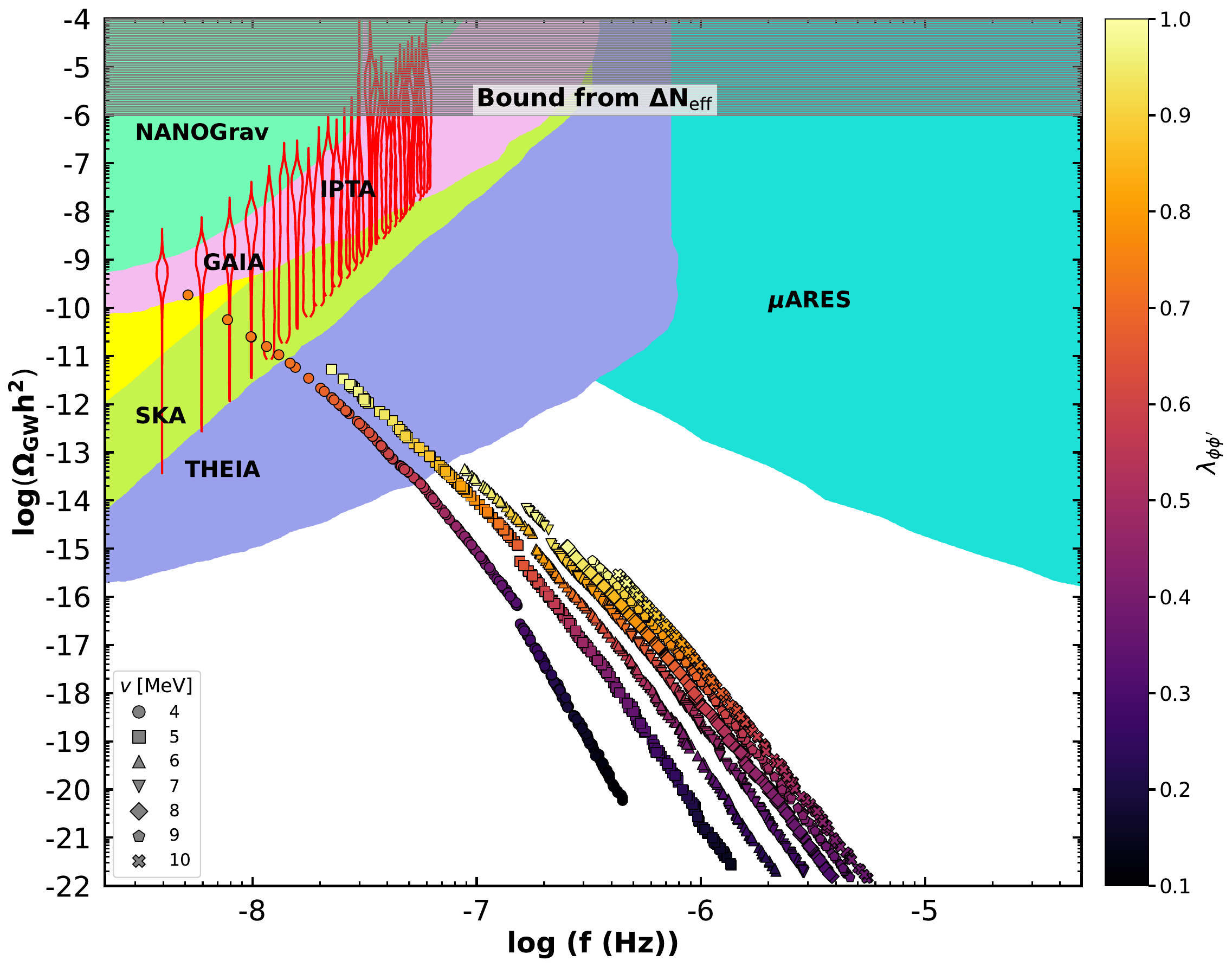}
\caption{{[\textit{Left}]: Results of the parameter scan in the $\beta/H_p$--$\alpha_p$ plane. The parameters $\lambda_\phi=0.01$, $\kappa_\phi=-0.05~\mathrm{MeV}$, and $\lambda_{\phi^\prime}=10^{-5}$ are held fixed, while the vacuum expectation value $v$ is varied from $4$ to $10~\mathrm{MeV}$ in discrete steps. The color scale denotes the value of the portal coupling $\lambda_{\phi\phi'}$. [\textit{Right}]: gravitational-wave spectra corresponding to the parameter points shown in the left panel, with different marker styles distinguishing the discrete values of the vacuum expectation value $v$.}}\label{fig:foptscan}
\end{figure*}

\begin{figure*}
    \centering   \includegraphics[scale=0.4]{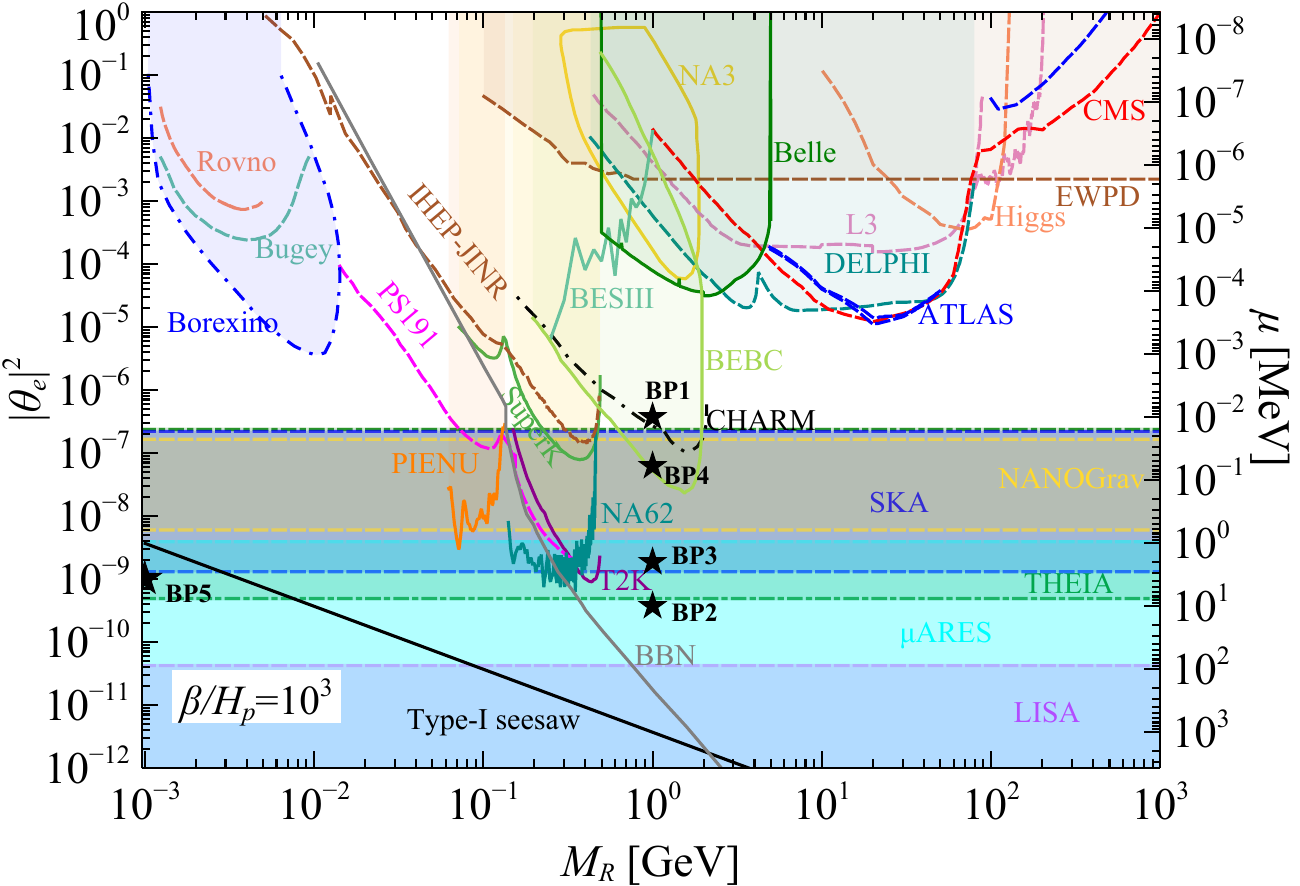}
    \includegraphics[scale=0.4]{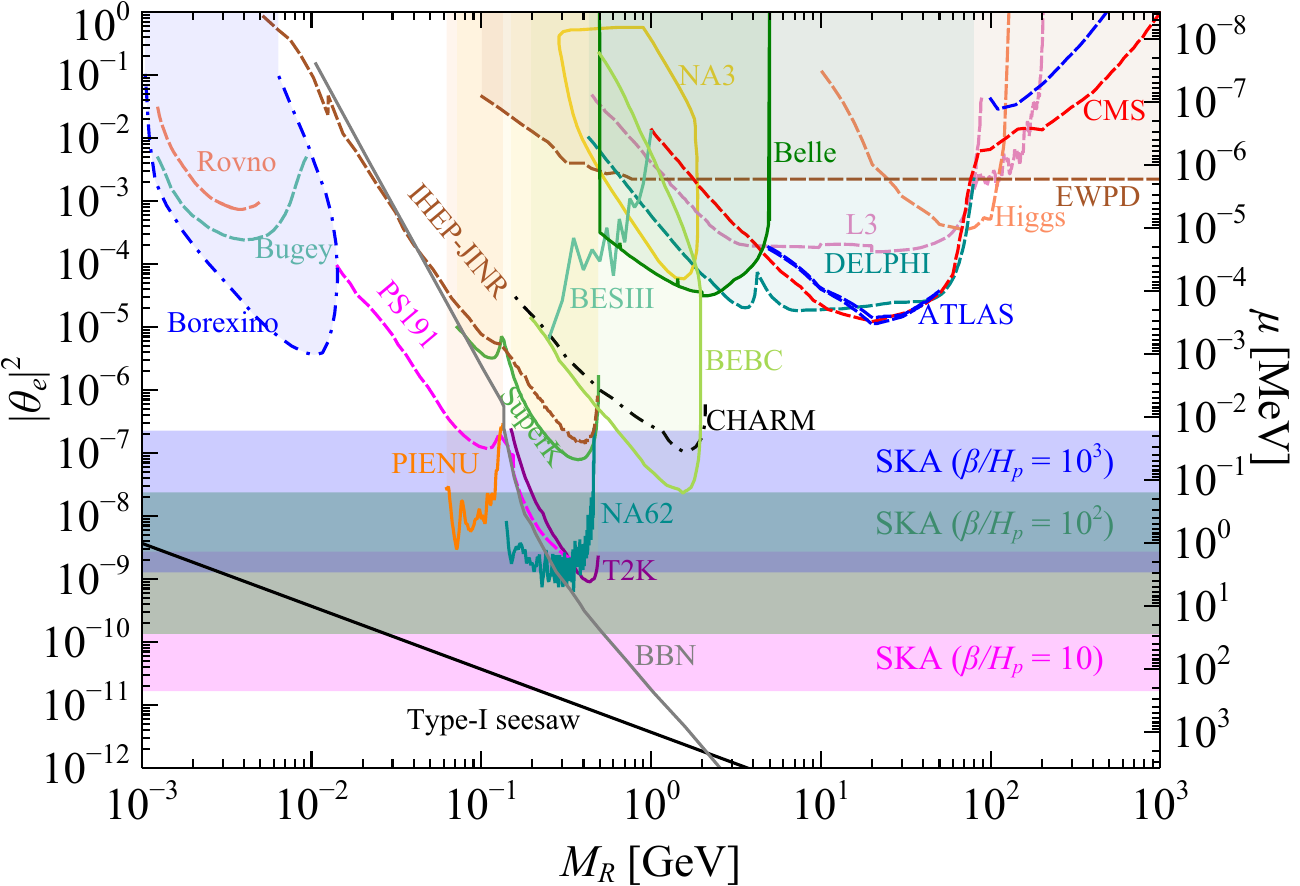}
    \caption{Exclusion limits on active-sterile mixing strength ($|\theta_e|^2\left[\equiv \sum_i |\Theta_{ei}|^2\right]$) as a function of $M_R$ for the inverse seesaw scenario. Different shaded contours show existing experimental exclusions \cite{Bolton:2022pyf} from collider searches (CMS \cite{CMS:2018iaf,CMS:2018jxx}, ATLAS \cite{ATLAS:2019kpx,ATLAS:2015gtp}, DELPHI \cite{DELPHI:1996qcc}, L3 \cite{L3:1992xaz,L3:2001zfe}, Belle \cite{Belle:2013ytx}), beam-dump and fixed-target experiments (CHARM \cite{CHARM:1985nku,CHARMII:1994jjr}, PS191 \cite{Bernardi:1987ek}, BEBC \cite{Barouki:2022bkt}, NA3 \cite{NA3:1986ahv}, NA62 \cite{NA62:2020mcv}, BESIII \cite{BESIII:2019oef}), electroweak precision data (EWPD \cite{Blennow:2016jkn,delAguila:2008pw,deBlas:2013gla,Antusch:2014woa}), Higgs searches \cite{Das:2017zjc}, and low-energy experiments (Borexino \cite{Borexino:2013bot}, Bugey \cite{Hagner:1995bn}, Rovno \cite{Derbin:1993wy}, PIENU \cite{PIENU:2017wbj}, T2K \cite{T2K:2019jwa}). The black diagonal line indicates the type-I seesaw prediction. The BBN \cite{Boyarsky:2009ix,Ruchayskiy:2012si} constraint is shown as a gray shaded region. (\textit{Left}) The projected sensitivities of future GW experiments SKA \cite{Weltman:2018zrl} (blue), THEIA \cite{Garcia-Bellido:2021zgu} (green), NANOGrav \cite{NANOGrav:2023gor,NANOGrav:2023ctt} (yellow), $\mu$ARES \cite{Sesana:2019vho} (cyan), and LISA \cite{LISA:2017pwj} (magenta) for a fixed $\beta/H_p=10^3$ are shown. (\textit{Right}) Dependence of the SKA \cite{Weltman:2018zrl} sensitivity bands on $\beta/H_p=10,10^2,10^3$, shown as horizontal colored bands.}
    \label{fig:summary1}
\end{figure*}

\section{Low scale FOPT}
The FOPT is driven by the singlet scalar $\phi$ as it rolls into its true vacuum and acquires a VEV. For simplicity, we ignore the couplings of $\phi, \phi'$ with the SM Higgs doublet. This keeps the low-scale FOPT decoupled from the electroweak sector as well as the TeV-scale heavy fermions $N_R, S_L$. We calculate the complete potential including the tree-level potential $V_{\rm tree}$ given in Eq. \eqref{eq:vtree}, one-loop Coleman-Weinberg potential $V_{\rm CW}$ \cite{Coleman:1973jx} along with the finite-temperature potential $V_{\rm th}$ \cite{Dolan:1973qd,Quiros:1999jp}, the details of which are given in Appendix \ref{app:fopt}.  The thermal field-dependent masses of different particles coupled to the singlet scalar $\phi$ are incorporated in the full potential. We then calculate the critical temperature $T_C$ at which the potential in the $\phi$ direction acquires another degenerate minima at $v_c = \phi (T=T_C)$. The order parameter of the FOPT is defined as $v_c/T_C$ such that a stronger FOPT corresponds to a larger $v_c/T_C$. The FOPT then proceeds via tunneling, the rate of which is estimated by calculating the bounce action $S_3$ using the prescription in \cite{Linde:1980tt, Guada:2020xnz}. The nucleation temperature $T_n$ is then calculated by comparing the tunneling rate with the Hubble expansion rate of the Universe: $\Gamma (T_n) = H^4(T_n)$. The transition is not complete when the first bubbles form. It is complete only when these bubbles have grown sufficiently to percolate throughout space, forming a connected cluster spanning the entire universe. It is this percolation moment that sets the physical timescale for both the completion of the phase transition and the production of the gravitational wave signal. We therefore use the percolation temperature $T_p$ as our reference scale, which is the temperature at which roughly 29\% of the universe has been converted to the true vacuum and a spanning cluster of bubbles first forms. On the other hand, it is the temperature at which a randomly chosen point has a 71\% chance of still being trapped in the false vacuum. Notably, $T_p$ is always lower than the conventional nucleation temperature $T_n$, with the gap becoming physically significant in the supercooled regime\footnote{See \cite{Athron:2023mer} for issues related to strong supercooled phase transition and explanation of the PTA data.}. We also estimate the reheat temperature $T_{\rm RH}$ at the end of the FOPT due to the release of radiation energy.

We then calculate the relevant parameters required to estimate the stochastic GW spectrum originating from the bubble collisions~\cite{Turner:1990rc,Kosowsky:1991ua,Kosowsky:1992rz,Kosowsky:1992vn,Turner:1992tz}, the sound wave of the plasma~\cite{Hindmarsh:2013xza,Giblin:2014qia,Hindmarsh:2015qta,Hindmarsh:2017gnf} and the turbulence of the plasma~\cite{Kamionkowski:1993fg,Kosowsky:2001xp,Caprini:2006jb,Gogoberidze:2007an,Caprini:2009yp,Niksa:2018ofa}. The two key parameters for GW estimate namely, the duration of the phase transition and the latent heat released relative to radiation density $(\rho_{\rm rad})$ are calculated and parametrized in terms of $\frac{\beta}{H_p}$ and $\alpha_p$ \cite{Caprini:2015zlo,Ellis:2018mja} respectively, where $H_p\equiv H(T_p)$. See Appendix \ref{app:fopt} for more details. 
\section{Results and Discussion}
Fig. \ref{fig:bp2} shows the potential profile at three different temperatures for one of the benchmark points (BP2) given in Table \ref{tabgw}. While the potential clearly has two degenerate minima at the critical temperature $T_C$, the profile remains identical at nucleation and percolation temperatures. The GW spectra for the benchmark points of Table \ref{tabgw} are shown in Fig. \ref{fig:gwspec}. The sensitivities of NANOGrav \cite{NANOGrav:2023ctt}, GAIA \cite{Garcia-Bellido:2021zgu}, THEIA~\cite{Garcia-Bellido:2021zgu}, $\mu$ARES~\cite{Sesana:2019vho}, IPTA~\cite{Manchester:2013ndt,Antoniadis:2022pcn}, and SKA~\cite{Weltman:2018zrl} are shown as different colored shades. The range of the GW spectrum from NANOGrav results~\cite{NANOGrav:2023gor} is shown by the red binned points.  The gray-shaded region above $\Omega_{\rm GW} h^2 \gtrsim 10^{-6}$ is disfavored by the limit on $\Delta N_{\rm eff}$ \cite{Planck:2018vyg} from GW overproduction. The yellow-shaded region is ruled out by cosmic microwave background (CMB) bounds \cite{Lasky:2015lej}. It is worth mentioning that MeV scale FOPT is tightly constrained from light nuclei synthesis during Big Bang Nucleosynthesis (BBN), restricting the reheat temperature $T_{\rm RH} \gtrsim 3$ MeV for strong FOPT $\alpha_p \gtrsim 1$ \cite{Bai:2021ibt}. This is satisfied by our benchmark points BP2, BP3, and BP4 with FOPT near the BBN epoch. On the other hand, limits on generating CMB distortions and excess of curvature perturbations can disfavor the benchmark points BP1 and BP2 \cite{Liu:2022lvz, Ramberg:2022irf, Elor:2023xbz, Xu:2025zsv}.

We now vary $v$ in the range 4 MeV to 10 MeV in discrete steps and $\lambda_{\phi\phi^\prime}$ in the range 0.1 to 1, while keeping the remaining parameters fixed at {$\lambda_\phi=0.01,\kappa_\phi=-0.05{~\rm MeV}, \lambda_{\phi^\prime}=10^{-5}$}. We keep only those parameter points that satisfy $v_c/T_C\gtrsim 1$ and display them in the \textit{left} panel of Fig. \ref{fig:foptscan} on the $\beta/H(T_p)-\alpha(T_p)$ plane. The color coding represents the values of $\lambda_{\phi\phi^\prime}$. We then calculate the peak amplitudes and peak frequencies for these parameter points and present them in the \textit{right} panel of Fig. \ref{fig:foptscan}, together with the projected sensitivities of current and future GW experiments. For a fixed VEV, as the portal coupling $\lambda_{\phi\phi^\prime}$ increases, the GW amplitude increases. This is because a larger $\lambda_{\phi\phi^\prime}$ strengthens the phase transition, leading to a larger $\alpha$ and a smaller $\beta/H$, thereby enhancing the GW signal, as can be easily seen from the \textit{left} panel of Fig. \ref{fig:foptscan}. An increase in the VEV shifts the FOPT to a higher energy scale, thereby increasing the peak GW frequency.

The experimental constraints and future projections for $|\theta_e|^2(\equiv \sum_i |\Theta_{ei}|^2)$ in the inverse seesaw are summarized in Fig.  \ref{fig:summary1}. The inverse seesaw scale $\mu$ is displayed on the right $y$-axis of the plots. We have taken the FOPT scale to be the same as the inverse seesaw scale, \textit{i.e.} $v=\mu$. In contrast to the type-I seesaw, the inverse seesaw naturally accommodates  $|\theta_e|^2$ well above the seesaw line, since neutrino mass suppression is achieved via a small lepton-number-violating Majorana mass $\mu\ll M_R$, decoupling the mixing angle from the neutrino mass scale. Current laboratory bounds exclude mixing angles above $|\theta_e|^2\sim10^{-9}-10^{-5}$ across $M_R\in[10^{-3},10^3]$ GeV, with T2K \cite{T2K:2019jwa}, NA62 \cite{NA62:2020mcv}, and PIENU \cite{PIENU:2017wbj} providing the strongest constraints at $M_R\lesssim1$ GeV, and ATLAS \cite{ATLAS:2019kpx,ATLAS:2015gtp}, CMS \cite{CMS:2018iaf,CMS:2018jxx}, and Belle \cite{Belle:2013ytx} dominating at higher masses. Future GW experiments SKA \cite{Weltman:2018zrl}, THEIA \cite{Garcia-Bellido:2021zgu}, $\mu$ARES \cite{Sesana:2019vho}, NANOGrav \cite{NANOGrav:2023ctt,NANOGrav:2023gor}, and LISA \cite{LISA:2017pwj} will extend sensitivity to regions of parameter space unreachable by collider or fixed-target searches, through the detection of the stochastic gravitational wave background sourced by a first-order phase transition, as shown in the \textit{left} panel. As shown in the \textit{right} panel, the GW (e.g., SKA \cite{Weltman:2018zrl}) reach is sensitive to the inverse phase transition parameter $\beta/H_p$: slower transitions produce a more energetic gravitational wave spectrum, yielding greater sensitivity at lower $|\theta_e|^2$.

In the \textit{left} panel of Fig. \ref{fig:summary1}, we have also shown the BPs from Table \ref{tabgw} with black star points. For the BP1, BP2, BP3 and BP4 we have fixed $M_R=1$ GeV, $m_1=0$, $R=I_{3\times3}$. The Yukawa coupling matrices satisfying the light neutrino mass and neutrino oscillation angles are given as
\begin{eqnarray}
 Y_D^{\rm BP1}=\begin{pmatrix}
     0&2.92&-1.62-i1.02\\
     0&3.49-i0.16&8.73\\
     0&-2.77-i0.17&9.26
\end{pmatrix}10^{-5}, \nonumber 
\end{eqnarray}
\begin{eqnarray}
 Y_D^{\rm BP2}=\begin{pmatrix}
     0&0.92&-0.51-i0.32\\
     0&1.1-i0.051&2.76\\
     0&-0.88-i0.054&2.93
\end{pmatrix}10^{-6}, \nonumber 
\end{eqnarray}
\begin{eqnarray}
 Y_D^{\rm BP3}=\begin{pmatrix}
     0&2.07&-1.15-i0.72\\
     0&2.47-i0.11&6.17\\
     0&-1.96-i0.12&6.55
\end{pmatrix}10^{-6}, \nonumber 
\end{eqnarray}
\begin{eqnarray}
 Y_D^{\rm BP4}=\begin{pmatrix}
     0&1.19&-0.66-i0.41\\
     0&1.42-i0.07&3.56\\
     0&-1.13-i0.07&3.78
\end{pmatrix}10^{-5}. \nonumber 
\end{eqnarray}

\section{Non-decoupling limit}
In the above analysis, the heavy singlet fermions $N_R, S_L$ are assumed to be decoupled from the low scale FOPT. However, for some region of the parameter space shown in Fig. \ref{fig:summary1}, the singlet fermion masses can be comparable to the FOPT scale and hence can not remain decoupled. Due to sizeable Yukawa interactions given by Eq. \eqref{eq:Ymu}, such light singlet fermions can significantly influence the phase transition dynamics. To account for this effect, we consider the non-decoupling limit by setting $Y_\mu = 1$, which corresponds to a seesaw scale of $\mu = v/\sqrt{2}$. The resulting GW spectrum for the benchmark point BP5, listed in Table~\ref{tabgw}, is shown in Fig.~\ref{fig:gwspec}. For this BP5, $M_R=1$ MeV, and we take two generations of $S_L$ and $N_R$. The predicted signal peaks within the sensitivity range of PTA experiments while remaining consistent with the bounds from BBN and curvature perturbations. For this BP5, the Yukawa coupling matrix satisfying the neutrino oscillation data is given as
\begin{eqnarray}
 Y_D^{\rm BP5}=\begin{pmatrix}
     0&1.56&-0.86-i0.54\\
     0&1.85-i0.09&4.64\\
     0&-1.47-i0.09&4.93
\end{pmatrix}10^{-10}. \nonumber 
\end{eqnarray}
 
\section{Conclusion}
We have studied the possibility of probing the dynamical origin of the inverse seesaw mechanism for light neutrino masses via observation of stochastic gravitational waves at pulsar timing array experiments. A scalar field is responsible for generating the lepton number-violating term of inverse seesaw dynamically by acquiring a non-zero VEV at a sub-MeV scale. Such a low-scale phase transition origin of the $\mu$ term in inverse seesaw not only keeps the heavy fermions around the TeV ballpark with large Yukawa couplings but also offers a complementary probe at GW experiments sensitive to low frequencies around nHz. We have considered a toy model in which a complex singlet scalar $\phi$ dynamically generates the lepton-number-violating term in the inverse seesaw. In order to ensure the first-order nature of the phase transition, we consider another scalar which couples to $\phi$ to generate a strong barrier between the degenerate minima of the scalar potential along the $\phi$ direction. We first identify a few benchmark points consistent with FOPT in the keV-MeV ballpark with corresponding GW spectra within reach of PTA, as well as other experiments like SKA, THEIA, and $\mu$ARES. We then project the corresponding parameter space in the plane of active-sterile mixing and heavy neutral lepton mass plane. While large active-sterile mixing can be probed by particle physics experiments, the parameter space corresponding to the smaller active-sterile mixing remains within reach of a variety of GW experiments. This offers interesting complementarity between particle physics and GW experiments strengthening the discovery prospects for dynamical inverse seesaw. It is worth mentioning that in the absence of any HNL signatures, a low-frequency GW signal alone can not be a definite probe of dynamical inverse seesaw. We also check the high scale validity of our toy model by solving the renormalization group (RG) evolution equations of different couplings, the details of which are given in Appendix \ref{app:rge}. While we have adopted a simplified toy model to perform our numerical calculations, richer UV completions are expected to keep the generic conclusions unchanged while offering additional degrees of freedom and detection aspects.
\section*{acknowledgments}
D.B. would like to acknowledge the hospitality at PITT-PACC, University of Pittsburgh during final stages of this work. P.K.P. acknowledges the Ministry of Education, Government of India, for providing financial support for his research via the Prime Minister’s Research Fellowship (PMRF) scheme. 

\appendix 
\section{UV complete realizations}
\label{appen1}
Here, we provide a UV complete model to realize the scenario discussed in the paper. The standard model particle content is extended by two singlet chiral fermions $S_L$ and $N_R$ along with two complex singlet scalars $\phi_1$ and $\phi$ to generate the non-zero neutrino mass in the inverse seesaw framework. Another complex singlet scalar $\phi^\prime$ is introduced to help $\phi$ obtain a strong first-order phase transition (FOPT). Additional discrete symmetries $Z_3$ and $Z_3^\prime$ are added to forbid the unwanted terms in the Lagrangian. The field content for minimal inverse seesaw realization is shown in Table \ref{tab1}.
\begin{table}[h]
    \centering
    \begin{tabular}{|c|c|}
    \hline
        Fields & $Z_3 \times Z'_3$ charge  \\
        \hline
        $\ell_L, \ell_R$ & $(\omega, 1)$ \\
        $N_R$ &   $(\omega, 1)$     \\
        $S_L$ & $(1, \omega)$ \\
        $\phi_1$ & $(\omega, \omega^2)$ \\
        $\phi$ & $ (1, \omega) $ \\
        $\phi^\prime$ & $ (1, 1) $ \\
         \hline 
    \end{tabular}
    \caption{Particles and their charge assignments under $Z_3\times Z_3^\prime$ symmetry.}
    \label{tab1}
\end{table}
The relevant part of the Lagrangian is 
\begin{equation}
    -\mathcal{L} \supset Y_D \overline{\ell_L} \tilde{\Phi} N_R + Y_R \overline{N_R} \phi_1 S_L + \frac{1}{2} Y_\mu \phi \overline{S^c_L} S_L + {\rm H.c.}
\end{equation}
The most general scalar potential is given as
\begin{eqnarray}
    V(\Phi,\phi_1,\phi,\phi^\prime)&=&-\mu_h^2 \Phi^\dagger \Phi+\lambda_h(\Phi^\dagger \Phi)^2-\mu_{\phi_1}^2\phi_1^\dagger\phi_1\nonumber\\&&+\lambda_{\phi_1}(\phi_1^\dagger\phi_1)^2-\mu_{\phi}^2\phi^\dagger\phi+\lambda_{\phi}(\phi^\dagger\phi)^2\nonumber\\&&+{\mu_{\phi^\prime}^2}\phi^{\prime\dagger}\phi^{\prime}+{\lambda_{\phi^\prime}}(\phi^{\prime\dagger}\phi^{\prime})^2\nonumber\\&&+\lambda_{\phi_1\phi}(\phi_1^\dagger\phi_1)(\phi^\dagger\phi)+\frac{\kappa_1}{3}(\phi_1^3+(\phi_1^\dagger)^3)\nonumber\\&&+\frac{\kappa_\phi}{3}(\phi^3+(\phi^\dagger)^3)+{\lambda_{\phi\phi^\prime}}(\phi^\dagger\phi)(\phi^{\prime\dagger}\phi^{\prime})\nonumber\\&&+{\lambda_{\phi_1\phi^\prime}}(\phi_1^\dagger\phi_1)(\phi^{\prime\dagger}\phi^{\prime}).
\end{eqnarray}
The quantum fluctuations around the minima of scalar fields $\Phi,\phi_1$, $\phi$ and \(\phi^\prime\) are given as
\begin{eqnarray}  \Phi&=&\left(0~~~\frac{h+v_{\rm ew}}{\sqrt{2}}\right)^T, \phi_1=\frac{\phi_1+v_{1}+i\eta_1}{\sqrt{2}},\nonumber\\& \phi=&\frac{\phi+v+i\eta}{\sqrt{2}},\phi^\prime=\frac{\phi^\prime+i\eta^\prime}{\sqrt{2}}.
\end{eqnarray}

After the scalar fields acquire non-zero vacuum expectation values (VEVs), we can write down the full neutral fermion mass matrix in the $(\nu_L, N_R, S_L)$ basis as 
\begin{equation}
    M_\nu = \begin{pmatrix}
        0 & M_D & 0 \\
        M^T_D & 0 & M_R \\
        0 & M^T_R & \mu \\
    \end{pmatrix},
\end{equation}
where $M_D = Y_D v_{\rm ew}/\sqrt{2}, M_R = Y_R v_1/\sqrt{2}, \mu = Y_\mu v/\sqrt{2}$ with $v_{\rm ew} =246$ GeV, $v_1, v$ being the VEVs of the standard model Higgs, singlet scalars $\phi_1$ and $\phi$ respectively. Similar realizations in which the lepton-number-violating term is generated dynamically through the spontaneous breaking of a gauged U(1) symmetry can be found in Refs. \cite{Ma:2009gu,Bazzocchi:2010dt, Khalil:2010iu, Baldes:2013eva, Ma:2015raa,DeRomeri:2017oxa, Borah:2021ocu,Panda:2022kbn,Berbig:2024uwm}.

\section{Parametrization of Yukawa coupling matrix}\label{app:yukawa}

The light neutrino mass matrix is given as
\begin{eqnarray}
m_\nu &=& -M_D M^{-1}_R \mu (M^T_R)^{-1} M^T_D,
\end{eqnarray}
which can be written as
\begin{eqnarray}
m_\nu= X\mu X^T,~~~{\rm with}~ X=M_D M_R^{-1}.\label{eq:muX}
\end{eqnarray}
The neutrino mass matrix $m_\nu$ can be diagonalized using the PMNS matrix $U_{\rm PMNS}$ and is related to the diagonal light neutrino mass matrix $m^d_\nu$ as
\begin{eqnarray}
    m_\nu=U^*_{\rm PMNS}.m_\nu^{\rm d}.U_{\rm PMNS}^T.\label{eq:mudiag}
\end{eqnarray}
Now using Eq \ref{eq:mudiag} in Eq \ref{eq:muX}, we get
\begin{eqnarray}
    X\mu X^T=U^*_{\rm PMNS}.m_\nu^{\rm d}.U_{\rm PMNS}^T,\nonumber\\U_{\rm PMNS}^T.X.\mu.X^T.U_{\rm PMNS}=m_\nu^d,\nonumber\\ Y\mu Y^T=m_\nu^d, ~~{\rm with~} Y=U_{\rm PMNS}^TX,\nonumber\\ (Y\sqrt{\mu})(Y\sqrt{\mu})^T=m_\nu^d,\nonumber\\ ZZ^T=m_\nu^d, ~~{\rm with~} Z=Y\sqrt{\mu}.\label{eq:Z1}
\end{eqnarray}
We can write
\begin{eqnarray}
    Z=\sqrt{m_\nu^d}R.\label{eq:Z2}
\end{eqnarray}
From Eq. \ref{eq:Z1} and Eq. \ref{eq:Z2} we have
\begin{eqnarray}
    \sqrt{m_\nu^d}R&=&Y\sqrt{\mu} \Rightarrow
Y=\sqrt{m_\nu^d}R(\sqrt{\mu})^{-1},\nonumber\\
U_{\rm PMNS}^T X&=&\sqrt{m_\nu^d}R(\sqrt{\mu})^{-1}\Rightarrow
X=U_{\rm PMNS}^*\sqrt{m_\nu^d}R (\sqrt{\mu})^{-1},\nonumber\\
    M_D M_R^{-1}&=&\frac{Y_Dv_{\rm ew}}{\sqrt{2}}M_R^{-1}=U_{\rm PMNS}^*\sqrt{m_\nu^d}R (\sqrt{\mu})^{-1}.
\end{eqnarray}
The Yukawa coupling matrix can then be expressed as
\begin{eqnarray}
    Y_D=\frac{\sqrt{2}}{v_{\rm ew}}U_{\rm PMNS}^*\sqrt{m_\nu^d}R (\sqrt{\mu})^{-1}M_R.
\end{eqnarray}
The active-sterile mixing is given as
\begin{eqnarray}
    \Theta\equiv M_DM_{R}^{-1}=U_{\rm PMNS}^*\sqrt{m_\nu^d}R (\sqrt{\mu})^{-1}.
\end{eqnarray}

\section{FOPT with the help of $\phi^\prime$}\label{app:fopt}

The finite temperature effective potential is given as
\begin{eqnarray}
    V_{\rm eff}(\phi,T)&=&V_{\rm tree}(\phi)+V_{\rm CW}(\phi)+V_{\rm ct}(\phi)+V_{\rm th}(\phi,T)\nonumber\\&&+V_{\rm daisy}(\phi,T),
\end{eqnarray}
where the tree level part of the potential is given as
\begin{eqnarray}
    V_{\rm tree}(\phi)=-\frac{1}{2}\left(\lambda_\phi v^2+\frac{\kappa_\phi v}{\sqrt{2}}\right)\phi^2+\frac{\lambda_\phi}{4}\phi^4+\frac{\kappa_\phi\phi^3}{3\sqrt{2}}.
\end{eqnarray}
The zero temperature one-loop Coleman-Weinberg potential in the $\overline{\rm MS}$ scheme is given by~\cite{Coleman:1973jx}
\begin{eqnarray}
    V_{\rm CW}(\phi)=\frac{1}{64\pi^2}\sum _in_im_i^4(\phi)\left( \log\left( \frac{m_i^2(\phi)}{\mu_R^2} \right)-c_i \right),\label{eq:cwpot}
\end{eqnarray}
where \(\mu_R^2\equiv v^2\) for our analysis, $c_i=3/2$ for scalar fields, and the degrees of freedom are given as
\begin{eqnarray}
    m_\phi^2(\phi)=-\mu_\phi^2+3\lambda_\phi \phi^2+\sqrt{2}\kappa_\phi\phi,~~~n_\phi=1,\\
    m_\eta^2(\phi)=-\mu_\phi^2+\lambda_\phi \phi^2-\sqrt{2}\kappa_\phi\phi,~~~n_\eta=1,\\
    m_{\phi^\prime}^2(\phi)=\mu_{\phi^\prime}^2+\frac{\lambda_{\phi\phi^\prime}}{2} \phi^2,~~~n_{\phi^\prime}=1,\\
    m_{\eta^\prime}^2(\phi)=\mu_{\phi^\prime}^2+\frac{\lambda_{\phi\phi^\prime}}{2} \phi^2,~~~n_{\eta^\prime}=1,
\end{eqnarray}
where $\mu_\phi^2=\lambda_\phi v^2+\frac{\kappa_\phi v}{\sqrt{2}}$.

The counter term can be written as
\begin{eqnarray}
    V_{\rm ct}(\phi)=-\frac{\delta\mu_\phi^2}{2}\phi^2+\frac{\delta\lambda_\phi}{4}\phi^4,
\end{eqnarray}
where the coefficients $\delta\mu_{\phi}^2$ and $\delta\lambda_\phi$ are obtained by solving the following Equations
\begin{eqnarray}
    \frac{\partial(V_{\rm CW}+V_{\rm ct})}{\partial\phi}\Bigg|_{\phi=v}=0,\frac{\partial^2(V_{\rm CW}+V_{\rm ct})}{\partial\phi^2}\Bigg|_{\phi=v}=0.
\end{eqnarray}
The thermal correction to the potential is given ~\cite{Dolan:1973qd,Quiros:1999jp}
\begin{eqnarray}
    V_{\rm th}(\phi,T)=\frac{T^4}{2\pi^2} \sum_jn_j J_B\left( \frac{m_j(\phi)}{T} \right),
\end{eqnarray}
where the thermal function is given as
\begin{eqnarray}
    J_{B}(y)=\int_0^\infty x^2\log\left( 1- e^{-\sqrt{x^2+y^2}} \right).
\end{eqnarray}
The daisy contribution is
\begin{eqnarray}
    V_{\rm daisy}=\frac{T}{12\pi}\sum_k\left( \left[m_k^2(\phi)\right]^{3/2} -\left[ m_k^2(\phi)+\Pi_k(T) \right]^{3/2} \right),\nonumber\\
\end{eqnarray}
where
\begin{eqnarray}
\Pi_\phi(T)=\Pi_\eta(T)=\left(\frac{\lambda_\phi}{3}+\frac{\lambda_{\phi\phi^\prime}}{12}\right)T^2,\nonumber\\
\Pi_{\phi^\prime}(T)=\Pi_{\eta^\prime}(T)=\left(\frac{\lambda_{\phi^\prime}}{3}+\frac{\lambda_{\phi\phi^\prime}}{12}\right)T^2.
\end{eqnarray}

The percolation temperature \(T_p\) is obtained from the probability that a point is still in false vacuum, given ~\cite{Guth:1981uk}
\begin{eqnarray}
    \label{C14}
    \mathcal{P}(T)=e^{-\mathcal{I}(T)},
\end{eqnarray}
where
\begin{eqnarray}
    \label{C15}
    \mathcal{I}(T)=\frac{4\pi}{3}\int_T^{T_C}\,\frac{\text{d}T'}{T'^4}\frac{\Gamma(T')}{H(T')}\left( \int_T^{T'}\frac{\mathrm{d}\tilde{T}}{H(\tilde{T})} \right)^3.
\end{eqnarray}
The percolation temperature is when \(\mathcal{I}(T_p)=0.34\). Further, the release of the latent heat as the phase transition completes will raise the temperature of the Universe, which we quantify with the reheating temperature by equating the free energy difference at the percolation temperature to the total energy density at that time, which can be expressed as~\cite{Ellis:2018mja}
\begin{eqnarray}
    \label{Treh}
    T_\mathrm{RH}=T_p\left(1+\alpha_p\right)^{1/4}
\end{eqnarray}
In order to calculate the energy released during the FOPT, we first find the free energy difference between the true and the false vacuum:
\begin{equation}
    \label{C17}
    \Delta V_\mathrm{tot}\equiv V_\mathrm{tot}\left( \phi_\mathrm{false},T \right) - V_\mathrm{tot}\left( \phi_\mathrm{true},T \right).
\end{equation}
As bubbles are nucleated, we can calculate the amount of vacuum energy released during the FOPT in terms of radiation energy density defined as~\cite{Ellis:2018mja,Athron:2023xlk}
\begin{equation}
    \label{C18}
    \alpha_p=\frac{\epsilon_p}{\rho_\mathrm{rad}}; \qquad \rho_\mathrm{rad}=\frac{\pi^2}{30}g_\ast T_p^4
\end{equation}
where
\begin{equation}
    \label{C19}
    \epsilon_p=\left[ \Delta V_\mathrm{tot} - \frac{T}{4}\frac{\partial \Delta V_{\rm tot}}{\partial T} \right]_{T=T_p}
\end{equation}
The energy released is also related to the change in the trace of the energy-momentum tensor across the bubble wall.
The duration of a FOPT is denoted by the parameter \(\beta\), and is defined as~\cite{Ellis:2018mja}
\begin{equation}
    \label{C20}
    \frac{\beta}{H(T)}\simeq T\frac{\mathrm{d}}{\mathrm{d}T}\left(\frac{S_3}{T}\right),
\end{equation}
and is calculated at \(T=T_p\). $S_3$ is the three-dimensional Euclidean action \cite{Linde:1980tt, Guada:2020xnz}.

\subsection{Non-decoupling limit}
In the non-decoupling limit, the field-dependent fermion mass terms are given as
\begin{eqnarray}
    m_i^+=\left( M_R^2+\frac{\mathcal{A}_i^2}{2}+ \mathcal{A}_i\sqrt{M_R^2+\frac{\mathcal{A}_i^2}{4}}\right)^{1/2},\\
    m_i^-=\left( M_R^2+\frac{\mathcal{A}_i^2}{2}- \mathcal{A}_i\sqrt{M_R^2+\frac{\mathcal{A}_i^2}{4}}\right)^{1/2},
\end{eqnarray}
where $\mathcal{A}_i=\frac{Y_{\mu_i}\phi}{\sqrt{2}}$, and $m_i^+,m_i^-$ are mass eigen states arising from $S_L$ and $N_R$ mixing.

\section{Renormalization Group Equations}\label{app:rge}
\subsection{Decoupling case}

\begin{figure}[h]
\centering
\includegraphics[scale=0.4]{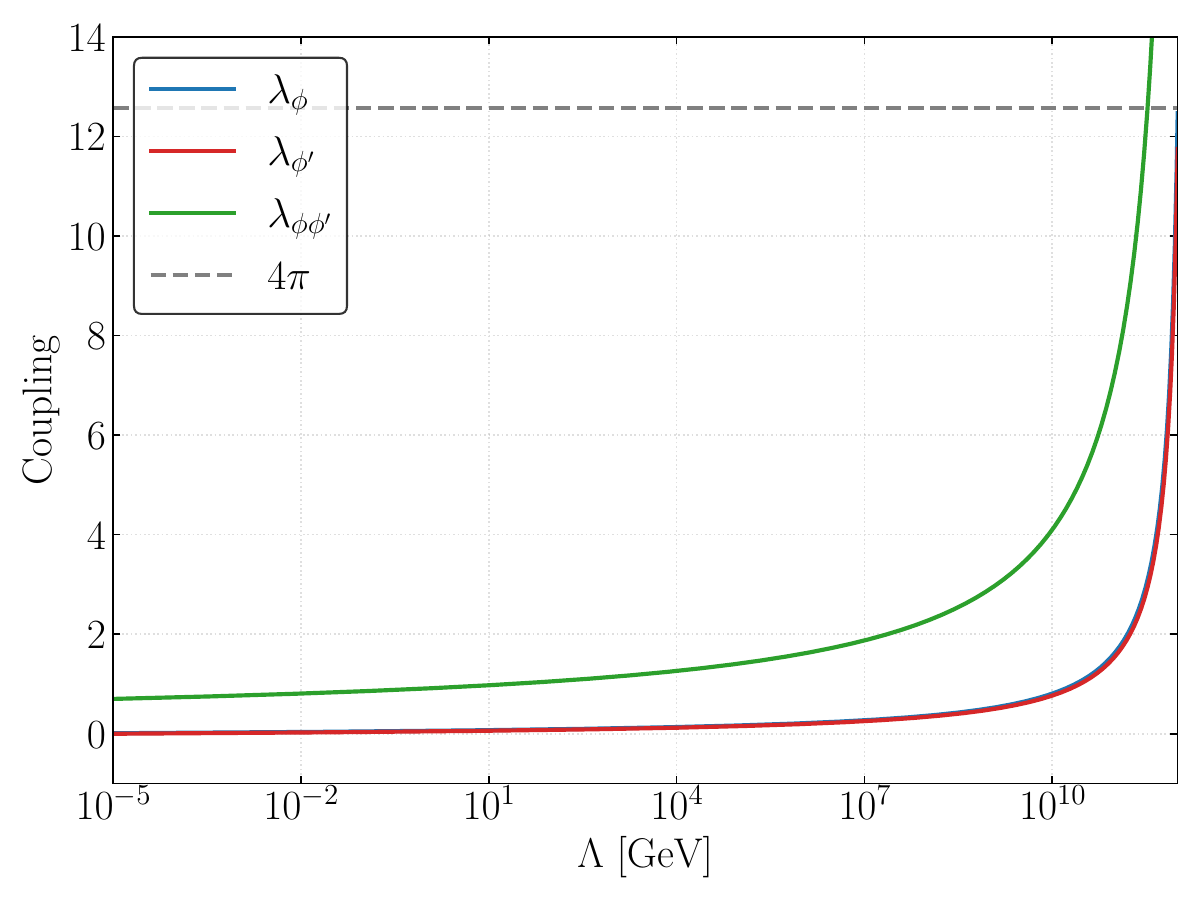}
\caption{Scalar couplings as function of $\Lambda$ for the BP1 as listed in Table \ref{tabgw}. The gray dashed line represents the coupling value $4\pi$.}
\label{fig:rgebp1}
\end{figure}

In the decoupling case, the fermions are assumed to be very large compared to the $\phi$ and hence do not enter into the RG evolution. The RG equations are given as
\begin{eqnarray}
\Lambda\frac{d\lambda_\phi}{d\Lambda}&=&\frac{1}{16\pi^2}(20\lambda_\phi^2+\lambda_{\phi\phi^\prime}^2),\\
\Lambda\frac{d\lambda_{\phi^\prime}}{d\Lambda}&=&\frac{1}{16\pi^2}(20\lambda_{\phi^\prime}^2+\lambda_{\phi\phi^\prime}^2),\\
\Lambda\frac{d\lambda_{\phi\phi^\prime}}{d\Lambda}&=&\frac{1}{16\pi^2}(8\lambda_\phi\lambda_{\phi\phi^\prime}+8\lambda_{\phi^\prime}\lambda_{\phi\phi^\prime}+4\lambda_{\phi\phi^\prime}^2),
\end{eqnarray}
where $\Lambda$ is the renormalization scale. In Fig. \ref{fig:rgebp1}, we show the RG running of the scalar couplings for the BP1 from Table \ref{tabgw}. We see that the portal coupling $\lambda_{\phi\phi^\prime}$ remains within the perturbative limit up to $\Lambda\sim10^{12}$ GeV.
\subsection{Non-decoupling case}
In this case, the fermions contribute to the RG equations. The RG equations are given as
\begin{eqnarray}
\Lambda\frac{d\lambda_\phi}{d\Lambda}&=&\frac{1}{16\pi^2}(20\lambda_\phi^2+\lambda_{\phi\phi^\prime}^2 +2\lambda_\phi Y_{\mu_1}^2 +2\lambda_\phi Y_{\mu_2}^2\nonumber\\&&- Y_{\mu_1}^4- Y_{\mu_2}^4 ),
\end{eqnarray}
\begin{eqnarray}
\Lambda\frac{d\lambda_{\phi^\prime}}{d\Lambda}&=&\frac{1}{16\pi^2}(20\lambda_{\phi^\prime}^2+\lambda_{\phi\phi^\prime}^2),
\end{eqnarray}
\begin{eqnarray}
\Lambda\frac{d\lambda_{\phi\phi^\prime}}{d\Lambda}&=&\frac{1}{16\pi^2}(8\lambda_\phi\lambda_{\phi\phi^\prime}+8\lambda_{\phi^\prime}\lambda_{\phi\phi^\prime}+4\lambda_{\phi\phi^\prime}^2\nonumber\\&&+\lambda_{\phi\phi^\prime}Y_{\mu_1}^2+\lambda_{\phi\phi^\prime}Y_{\mu_2}^2),
\end{eqnarray}
\begin{eqnarray}
\Lambda\frac{d Y_{\mu_1}}{d\Lambda}&=&\frac{1}{16\pi^2}\left(\frac{3}{2}Y_{\mu_1}^3+\frac{1}{2}Y_{\mu_1}Y_{\mu_2}^2\right),
\end{eqnarray}
\begin{eqnarray}
\Lambda\frac{d Y_{\mu_2}}{d\Lambda}&=&\frac{1}{16\pi^2}\left(\frac{3}{2}Y_{\mu_2}^3+\frac{1}{2}Y_{\mu_2}Y_{\mu_1}^2\right),
\end{eqnarray}
\begin{figure}[h]
\centering
\includegraphics[scale=0.4]{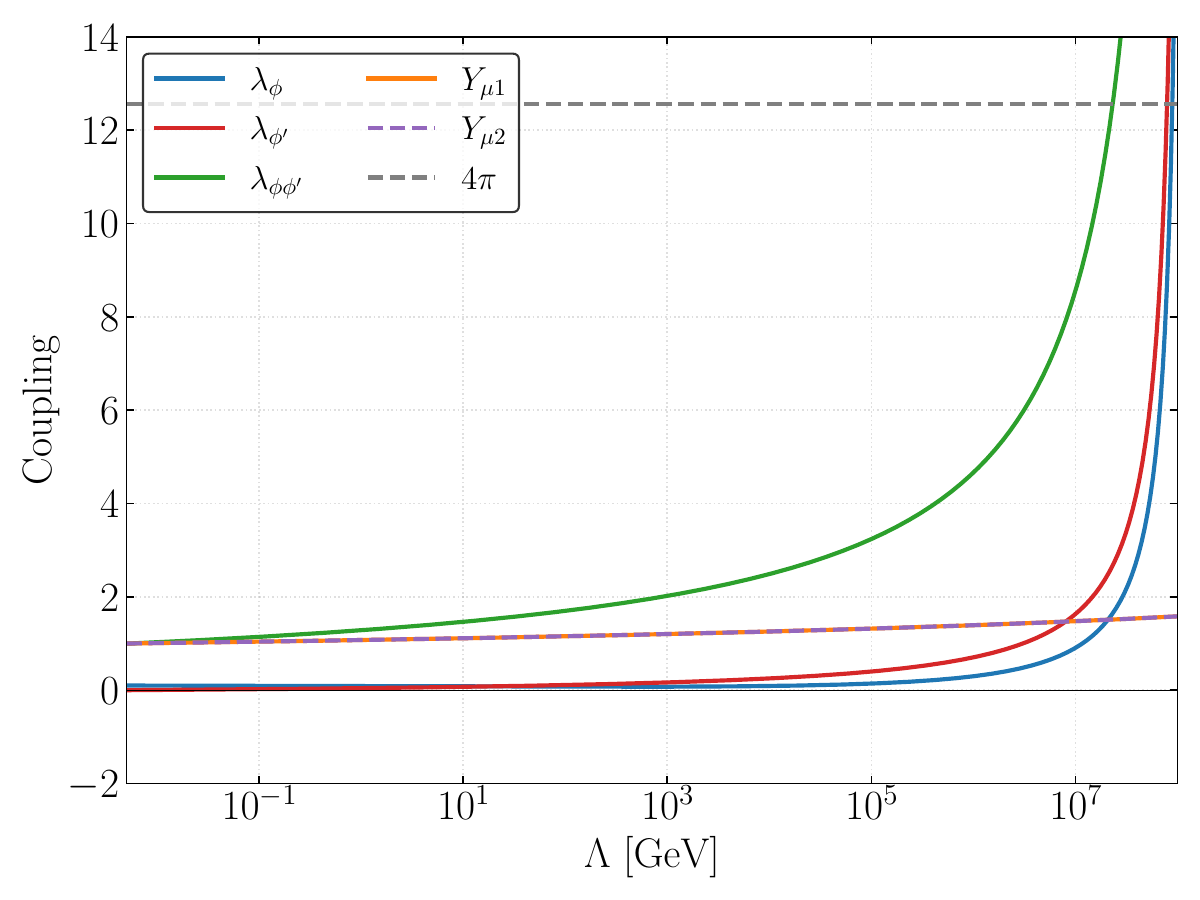}
\caption{Scalar couplings and Yukawa couplings as function of $\Lambda$ for the BP5 as listed in Table \ref{tabgw}. The gray dashed line represents the coupling value $4\pi$.}
\label{fig:rgebp5}
\end{figure}
In Fig. \ref{fig:rgebp5}, we show the RG running of the scalar couplings and Yukawa couplings for the BP5 from Table \ref{tabgw}. The Yukawa couplings are taken to be $Y_{\mu_1}=Y_{\mu_2}=1$. We see that the portal coupling $\lambda_{\phi\phi^\prime}$ remains within the perturbative limit up to $\Lambda\sim10^7$ GeV.


\providecommand{\href}[2]{#2}\begingroup\raggedright\endgroup

\end{document}